\documentstyle[aps,twocolumn,prl,epsfig]{revtex}

\begin{document}

\twocolumn[\hsize\textwidth\columnwidth\hsize\csname
@twocolumnfalse\endcsname \draft
\title{Single hole motion in LaMnO$_{3}$}
\author{Wei-Guo Yin$^{1,2}$, Hai-Qing Lin$^{1}$, and Chang-De Gong$^{2}$}
\address{$^1$Department of Physics, The Chinese University of Hong Kong, Sha Tin,
Hong Kong, People's Republic of China}
\address{$^2$National Key Laboratory of Solid States of Microstructure, Nanjing
University, Nanjing 210093, People's Republic of China}
\date{\today}
\maketitle

\begin{abstract}
We study single hole motion in LaMnO$_{3}$ using the classical
approximation for JT lattice distortions, a modified Lang-Firsov
approximation for dynamical breathing-mode phonons, and the
self-consistent Born approximation (verified by exact
diagonalization) for hole-orbital-excitation scattering. We show
that in the realistic parameter space for LaMnO$_{3}$, quantum
effects of electron-phonon interaction are small. The
quasiparticle bandwidth $W\simeq 2.2J$ in the purely orbital
$t$-$J$ model. It is strikingly broadened to be of order $t$ by
strong static Jahn-Teller lattice distortions even when the
polaronic band narrowing is taken into account.
\end{abstract}


]

LaMnO$_{3}$ is a typical parent compound of a class of colossal
magnetoresistance materials $R_{1-x}A_{x}$MnO$_{3}$ ($R$ = La, Pr, Nd, Sm
and $A$ = Ca, Sr, Ba). As temperature decreases, it undergoes a structural
phase transition at $T_{s}\sim 750$ K from cubic to tetragonal because of
Jahn-Teller (JT) lattice distortions of its MnO$_{6}$ octahedra. Below $%
T_{N}\sim 140$ K, it is an $A$-type antiferromagnetic (AF) insulator.
Staggered orbital ordering was recently observed in the ferromagentic planes
of LaMnO$_{3}$ \cite{Murakami}, which can be driven by either the
intra-atomic Coulomb interaction in the $e_{g}$ orbitals \cite{Nagaosa,Brink}
or the cooperative JT splitting of the degenerate $e_{g}$ orbitals \cite
{Kanamori,Hotta}. When electrons are removed from the $e_{g}$ orbitals upon
doping, charge fluctuations as well as spin, orbital defects are introduced
into the system. LaMnO$_{3}$ with one hole is one of the simplest real
systems containing charge, spin, orbital, and lattice dynamics. The clear
understanding of single hole motion in it is an essential step to the full
understanding of abundant dopant-induced phase transition phenomena observed
in this class of materials \cite{Tokura}.

The Coulomb interaction ($\sim 7$ eV) in the $e_{g}$ orbitals eliminates
doubly occupied sites and results in a two-dimensional anisotropic orbital $t
$-$J$ model \cite{Kugel}. There exist off-diagonal transfer matrix elements,
thus in principle a hole can hop without disturbing the staggered orbital
background. If the orbitals were frozen, a hole would move freely and
disperse with bandwidth $2t$, as described by the large-$U$ LDA$+U$ band
calculation \cite{Satpathy}. However, the low-energy physics may be
controlled by two polarization effects: {\em orbital polarization} and {\em %
lattice polarization}. First, the motion of a hole will distort the orbital
order via its diagonal transfer matrix elements; this will induce a strong
polarized cloud of orbital excitations in the vicinity of the hole. The
resultant quasiparticle (QP) is of orbital-polaron type with bandwidth $%
\propto J$ and a shift of spectral weight from the coherent to the
incoherent part of the hole spectrum \cite{Kilian}. Second, when a
hole is present, it attracts the surrounding oxygen ions equally,
giving rise to a breathing distortion energy. In the single hole
problem, charge fluctuations exactly locate at where the hole is
and thus accompany the hole propagation. One thus needs to take
into account {\em dynamical} breathing-mode phonons. Beyond a
critical hole-phonon coupling strength, the hole is self-trapped
in an ``anti-JT'' small polaron state via the polaronic band
narrowing effect \cite{Allen,Millis,Roder,note1}. Therefore, the
QP behavior is sensitive to model parameters. In this Letter, we
present a systematic study on single hole motion in LaMnO$_3$ in a
variety of parameter regions of physical interest.

The total Hamiltonian consider here is $H=H_{t-J}+H_{{\rm br}}+H_{{\rm JT}}$
\cite{Hotta,Kugel,Millis,Roder}, where
\begin{eqnarray}
H_{t-J} &=&-\sum_{\langle {\bf ij}\rangle \parallel ab}(t_{{\bf ij}}^{ab}%
\widetilde{d}\,_{{\bf i}a}^{\dagger }\widetilde{d}\,_{{\bf j}b}^{{}}+{\rm %
H.c.})  \nonumber \\
&&+\frac{J}{2}\sum_{\langle {\bf ij}\rangle \parallel }[T_{{\bf i}}^{z}T_{%
{\bf j}}^{z}+3T_{{\bf i}}^{x}T_{{\bf j}}^{x}\mp \sqrt{3}(T_{{\bf i}}^{x}T_{%
{\bf j}}^{z}+T_{{\bf i}}^{z}T_{{\bf j}}^{x})],  \label{model} \\
H_{{\rm br}} &=&-\sqrt{E_{{\rm br}}\omega _{0\,}}\sum_{{\bf i}}n_{{\bf i}%
}^{h}(a_{{\bf i}}^{\dagger }+a_{{\bf i}}^{{}})+\omega _{0}\sum_{{\bf i}}a_{%
{\bf i}}^{\dagger }a_{{\bf i}}^{{}},  \nonumber \\
H_{{\rm JT}} &=&2\lambda \sum_{{\bf i}}(Q_{2{\bf i}}T_{{\bf i}}^{z}+Q_{3{\bf %
i}}T_{{\bf i}}^{x})+\frac{1}{2}K\sum_{{\bf i}}(Q_{2{\bf i}}^{2}+Q_{3{\bf i}%
}^{2}),  \nonumber
\end{eqnarray}
where $\widetilde{d}\,_{{\bf i}a}^{\dagger }=d\,_{{\bf i}a}^{\dagger }(1-n_{%
{\bf i}\overline{a}})$ is the constrained fermion operator for the $e_{g}$
electron at orbital $a$. $T_{{\bf i}}^{z}=(\widetilde{d}\,_{{\bf i}\uparrow
}^{\dagger }\widetilde{d}\,_{{\bf i}\uparrow }^{{}}-\widetilde{d}\,_{{\bf i}%
\downarrow }^{\dagger }\widetilde{d}\,_{{\bf i}\downarrow }^{{}})/2$ and $T_{%
{\bf i}}^{x}=(\widetilde{d}\,_{{\bf i}\uparrow }^{\dagger }\widetilde{d}\,_{%
{\bf i}\downarrow }^{{}}+\widetilde{d}\,_{{\bf i}\downarrow }^{\dagger }%
\widetilde{d}\,_{{\bf i}\uparrow }^{{}})/2$ are orbital-pseudospin operators
with $|\uparrow \rangle =d_{x^{2}-y^{2}}$ and $|\downarrow \rangle
=d_{3z^{2}-r^{2}}$. The anisotropic transfer matrix elements are $t_{{\bf ij}%
}^{\uparrow \uparrow }=3t/4$, $t_{{\bf ij}}^{\downarrow \downarrow
}=t/4$, and $t_{{\bf ij}}^{\uparrow \downarrow }=t_{{\bf
ij}}^{\downarrow \uparrow }=\mp \sqrt{3}t/4$, here the $\mp $ sign
distinguishes hopping along the $x$ and $y$ directions. The
orbital superexchange $J$-induced orbital ordering reproduces the
experimental observation: $(|\uparrow \rangle +|\downarrow \rangle
)/\sqrt{2}$ and $(|\uparrow \rangle -|\downarrow \rangle
)/\sqrt{2}$ in the $A$ and $B$ sublattices respectively
\cite{Murakami,Brink}. In $H_{{\rm br}}$, $n_{{\bf i}}^{h}=1-\sum_{a}%
\widetilde{d}\,_{{\bf i}a}^{\dagger }\widetilde{d}\,_{{\bf i}a}$
is the hole number operator. $a_{{\bf i}}$'s are the
breathing-mode phonon operators with frequency $\omega _{0}$.
$E_{{\rm br}}$ is the hole-phonon
coupling strength$.$ In $H_{{\rm JT}}$ \cite{note1}, $Q_{2{\bf i}}$ and $Q_{3%
{\bf i}}$ are, respectively, JT distortions for the $3z^{2}-r^{2}$ and $%
x^{2}-y^{2}$ modes satisfying $Q_{2{\bf i}}=q\cos \psi _{{\bf i}}$ and $Q_{3%
{\bf i}}=q\sin \psi _{{\bf i}}$ with $\psi _{{\bf i}\in A}=-\psi _{{\bf j}%
\in B}$ \cite{Matsumoto}. For simplicity, $\psi _{{\bf i}\in A}$ is set to
be $\pi /2$ so that the orbital ordering driven by the cooperative JT effect
also agrees with the experiment.

We employ the slave-fermion formalism to cope with the constraint of no
doubly occupancy \cite{SVR}. First, we perform a uniform rotation of
orbitals by $90^{\circ }$ about the $T^{y}$ axis: $\widetilde{d}_{{\bf i}%
\uparrow }\rightarrow (\widetilde{c}_{{\bf i}\uparrow }-\widetilde{c}_{{\bf i%
}\downarrow })/\sqrt{2}$, $\widetilde{d}_{{\bf i}\downarrow }\rightarrow (%
\widetilde{c}_{{\bf i}\uparrow }+\widetilde{c}_{{\bf i}\downarrow })/\sqrt{2}
$,$\,T_{{\bf i}}^{z}\rightarrow -\widetilde{T}\,_{{\bf i}}^{x}$, $T_{{\bf i}%
}^{x}\rightarrow \widetilde{T}\,_{{\bf i}}^{z}$ in order to obtain the
N\'{e}el configuration $|\cdots \widetilde{T}\,_{{\bf i}}^{z}\widetilde{T}%
\,_{{\bf i}+1}^{z}\widetilde{T}\,_{{\bf i}+2}^{z}\widetilde{T}\,_{{\bf i}%
+3}^{z}\cdots \rangle =|\cdots \downarrow \uparrow \downarrow \uparrow
\cdots \rangle $. Then, considering the N\'{e}el state as the vacuum state,
we define holon (spinless fermion) operators $h_{{\bf i}}$ so that $%
\widetilde{c}_{{\bf i}\uparrow }=h\,_{{\bf i}}^{\dagger }b_{{\bf i}%
}^{{}},\,\,\widetilde{\,c}_{{\bf i}\downarrow }=h_{{\bf i}}^{\dagger }$ on
the $\downarrow $ sublattice and $\widetilde{\,c}_{{\bf j}\downarrow }=h_{%
{\bf j}}^{\dagger }b_{{\bf j}}^{{}},\,\,\,\widetilde{c}_{{\bf j}\uparrow
}=h_{{\bf j}}^{\dagger }$ on the $\uparrow $ sublattice. Here $b_{{\bf i}}=%
\widetilde{T}\,_{{\bf i}}^{-}$ on the $\downarrow $ sublattice and $%
\widetilde{T}\,_{{\bf i}}^{+}$ on the $\uparrow $ sublattice are hard-core
boson operators. Such a treatment is in spirit similar to that done by
Schmitt-Rink {\em et al.} for one hole motion in a quantum antiferromagnet
\cite{SVR,MH,Yin}.

Following R\"{o}der, Zang, and Bishop \cite{Roder}, we treat dynamical
phonons within the modified Lang-Firsov approximation using a canonical
transformation $\overline{H}=U^{\dagger }HU$, $U=e^{-S_{1}(\Delta
)}e^{-S_{2}(\tau )}$, where $S_{1}(\Delta )=\Delta /2\sqrt{E_{{\rm br}%
}\omega _{0}}\sum_{{\bf i}}(a_{{\bf i}}^{\dagger }-a_{{\bf i}}^{{}})$ with $%
\Delta $ measuring static breathing-mode distortions and $S_{2}(\tau )=-\tau
\sqrt{E_{{\rm br}}/\omega _{0}}\sum_{i}h\,_{{\bf i}}^{\dagger }h_{{\bf i}%
}^{{}}(a_{{\bf i}}^{\dagger }-a_{{\bf i}}^{{}})$ with $\tau $ measuring the
degree of the polaron effect. Averaging $\overline{H}$ over the phonon
vacuum, we arrive at an effective orbital-lattice-polaron Hamiltonian $%
\overline{H}=E_{0}(\tau )+H_{{\rm eff}}$
\begin{eqnarray}
H_{{\rm eff}} &=&\sum_{{\bf k}}\varepsilon _{{\bf k}}(\tau )h\,_{{\bf k}%
}^{\dagger }h_{{\bf k}}^{{}}+\sum_{{\bf q}}\omega _{{\bf q}}\beta _{{\bf q}%
}^{\dagger }\beta _{{\bf q}}^{{}}  \nonumber \\
&&+\sum_{{\bf kq}}h\,_{{\bf k}}^{\dagger }h_{{\bf k}-{\bf q}}^{{}}(M_{{\bf kq%
}}^{{}}\beta _{{\bf q}}^{{}}+N_{{\bf kq}}^{{}}\beta _{{\bf q+Q}}^{{}})+{\rm %
H.c.}
\end{eqnarray}
where ${\bf Q=(}\pi ,\pi {\bf )}$. $E_{0}(\tau )=Nx[(1-\tau )\Delta +\Delta
^{2}/4E_{{\rm br}}-E_{{\rm br}}(2\tau -\tau ^{2})]$ with $x=1/N$ being the
hole concentration and $\Delta =2E_{{\rm br}}(1-\tau )x$. The
hole-orbital-wave coupling functions are $M_{{\bf kq}}=\frac{2t}{\sqrt{N}}%
(\gamma _{{\bf k}}v_{{\bf q}}+\gamma _{{\bf k}-{\bf q}}u_{{\bf q}})$, $N_{%
{\bf kq}}=-\frac{\sqrt{3}t}{\sqrt{N}}(\eta _{{\bf k}}v_{{\bf q}}-\eta _{{\bf %
k}-{\bf q}}u_{{\bf q}})$ with $u_{{\bf q}}=\{[(A_{{\bf q}}+B_{{\bf q}%
})/\omega _{{\bf q}}+1]/2\}^{1/2}$ and $v_{{\bf q}}=-{\rm sgn}(B_{{\bf q}%
})\{[(A_{{\bf q}}+B_{{\bf q}})/\omega _{{\bf q}}-1]/2\}^{1/2}$. Here short
notations are $\gamma _{{\bf k}}=(\cos k_{x}+\cos k_{y})/2$, $\eta _{{\bf k}%
}=(\cos k_{x}-\cos k_{y})/2$, $A_{{\bf q}}=3${\bf $J+$}$E_{{\rm JT}}$, and $%
B_{{\bf q}}=${\bf $J$}$\gamma _{{\bf q}}/2$. The $\beta _{{\bf q}}$'s are
orbital wave operators, $b_{{\bf q}}=u_{{\bf q}}\beta _{{\bf q}}+v_{{\bf q}%
}\beta _{-{\bf q}}^{\dagger }$, $\,$with dispersion $\omega _{{\bf q}}=\sqrt{%
A_{{\bf q}}(A_{{\bf q}}+2B_{{\bf q}})}$. The bare hole dispersion is $%
\varepsilon _{{\bf k}}(\tau )=-\xi (\tau )t\gamma _{{\bf k}}$ with $\xi
(\tau )=\exp (-E_{{\rm br}}\tau ^{2}/\omega _{0})$ being the polaronic band
narrowing. $\xi (\tau )$ is determined by the following equation \cite{Roder}%
:

\begin{equation}
E_{{\rm br}}=-\omega _{0}\ln \xi \left[ 1-\frac{1}{(1+x)\omega _{0}}\frac{%
\partial E_{\min }(\xi )}{\partial \xi }\right] ^{2},
\end{equation}
where $E_{\min }(\xi )$ is the minimum of the QP dispersion $E_{{\bf k}%
}\equiv \varepsilon _{{\bf k}}+{\rm Re}\Sigma ({\bf k},E_{{\bf k}})$ with $%
\Sigma ({\bf k},\omega )$ being the holon self-energy. Note that in this
effective Hamiltonian, the JT effect acts as $E_{{\rm JT}}=2\lambda ^{2}/K$
in $A_{{\bf q}}$.

We calculate the holon Green's function $G({\bf k},\omega
)=[\omega -\varepsilon _{{\bf k}}-\Sigma ({\bf k},\omega )]^{-1}$
treating the hole-orbital-wave coupling within the self-consistent
Born approximation (SCBA) \cite{SVR,MH,Yin}. The self-energy is
thus of the form
\begin{figure}[t]
\begin{center}
\vskip 0cm \epsfig{file=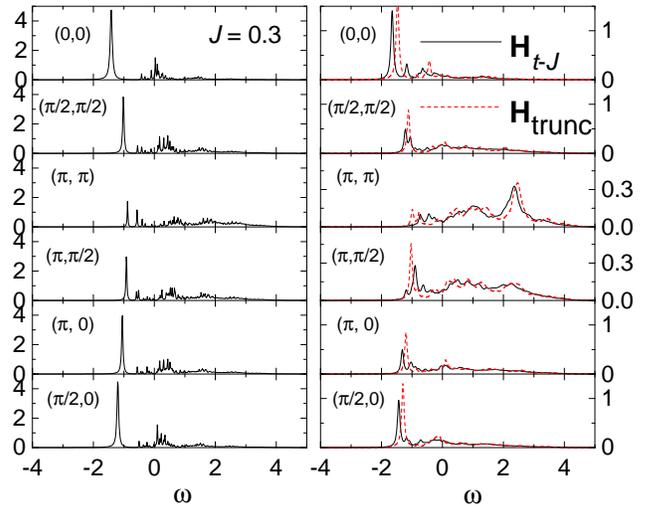, width=\hsize}
\end{center}
\caption{QP spectral functions $A({\bf k},\omega )=-{\rm Im}G({\bf
k},\omega )/\pi$ for the orbital $t$-$J$ model on the $4\times 4$
cluster calculated using the SCBA (left panel) and using the ED
technique (right panel).} \label{fig. 1}
\end{figure}
\begin{figure}[t]
\begin{center}
\vskip 0.1cm \epsfig{file=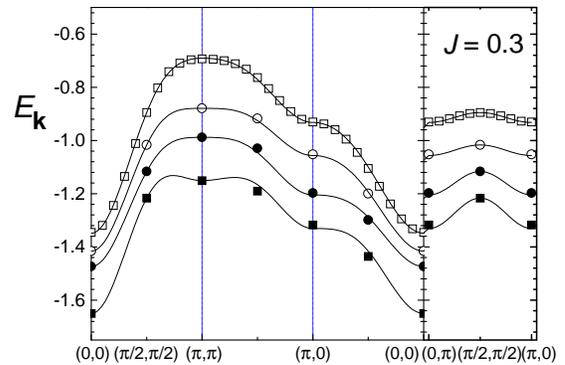, width=\hsize}
\end{center}
\caption{QP dispersion for the orbital $t$-$J$ model with
$J=0.3t$: The ED results for the full Hamiltonian (solid squares)
and the truncated Hamiltonian (solid circles); The SCBA results on
the $4\times 4$ cluster (open circles) and the $20\times 20$ one
(open squares). Solid lines are the fits using $E_{{\bf
k}}^{f}=a_{0}+a_{1}\gamma _{{\bf k}}+a_{2}\cos k_{x}\cos
k_{y}+a_{3}\gamma _{2{\bf k}}$.} \label{fig. 2} \vskip 0.4cm
\end{figure}
\begin{eqnarray}
\Sigma ({\bf k},\omega ) &=&\sum_{{\bf q}}{}[M_{{\bf kq}}^{2}G({\bf k}-{\bf q%
},\omega -\omega _{{\bf q}})  \nonumber \\
&&+N_{{\bf kq}}^{2}G({\bf k}-{\bf q},\omega -\omega _{{\bf q+Q}})].
\label{se}
\end{eqnarray}
We test the applicability of the SCBA by numerically diagonalizing $H_{t%
\text{-}J}$ on $16$- and $18$-site clusters. In addition, since in the SCBA
the AF orbital background is treated within linear orbital-wave theory, the
mixed terms $\propto T_{{\bf i}}^{x}T_{{\bf j}}^{z}+T_{{\bf i}}^{z}T_{{\bf j}%
}^{x}$ in $H_{t\text{-}J}$, which contribute only in higher order
orbital-wave theory, are neglected in the SCBA. It is thus interesting to
numerically diagonalize a ``truncated'' Hamiltonian without these terms \cite
{Brink}. As shown in {\rm Fig. 1 and Fig. 2}, we find that all of the SCBA
results are in good agreement with the exact diagonalization (ED)\ results,
especially with those for the truncated Hamiltonian as expected.

The physically relevant parameter space for LaMnO$_{3}$ are:
$t\sim 0.72$~eV (energy unit) and $0.1\leq J\leq 0.3$ are
estimated from photoemission experiments \cite{Saitoh}. $\omega
_{0}=0.1$ is taken from Raman data \cite {Iliev}. $E_{{\rm
JT}}=1.5$ and $E_{{\rm br}}$ is of order $E_{{\rm JT}}$
\cite{Kanamori,Hotta,Millis,Roder}. Numerical studies found that the $A$%
-type spin and $C$-type orbital structures were stabilized in this region of
parameter space \cite{Hotta}. Below let us discuss the QP properties in
different parameter regions of interest. All calculations are performed on
the $16 \times 16$ lattice unless noted.

First, we consider the single hole motion in the pure orbital
$t$-$J$ model (i.e. $E_{{\rm br}}=E_{{\rm JT}}=0$).
As displayed in {\rm Fig. 1}, at any $%
{\bf k}$, there is a well-defined quasiparticle pole (i.e., zero
orbital-wave) at the low energy side which is well separated from a broad,
incoherent, multiple-orbital-wave background extending to the full
free-electron bandwidth. The bottom and the top of the quasiparticle (QP)
band locate at $(0,0)$ and $(\pi ,\pi )$, respectively. In Fig. 2, the
spectrum of the orbital polaron is flat at large momenta, which leads to a
strongly distorted density of states with a massive peak at the top of the
QP band. In the realistic range of $0.01\leq J\leq 0.5$, the QP bandwidth $%
W\simeq 2.2J$ scales with $J$ (see {\rm Fig. 3}). This new low
energy scale is quite similar to that in the cuprate $t$-$J$ model
where the spin is conserved during hole hopping \cite{SVR,MH,Yin}.
In the latter, the staggered spin background is disturbed by hole
propagation, leading to a vanishing bare hole dispersion
$\varepsilon _{{\bf k}}$, and thereafter restored by Heisenberg
spin flipping, forming a QP band with width $\sim 2J$
(the characteristic energy of spin waves). In the present case of $%
\varepsilon _{{\bf k}}=-t\gamma _{{\bf k}}$, the energy scale of
$W\simeq 2.2J $ can be understood in the following way: For small
$J$, orbital excitations are easily stimulated by incoherent hole
motion and accompany the hole
propagation. Thus the hole spectral weights $Z({\bf k})=[1-\partial \Sigma (%
{\bf k},\omega )/\partial \omega ]_{\omega =E_{{\bf k}}}^{-1}$ are strongly
reduced by such a cloud of polarized orbital waves. The reduced weights by
incoherent hole motion can be approximately obtained by neglecting $%
\varepsilon _{{\bf k}}$ in the calculation, referred to as $Z_{0}({\bf k})$.
Then one can naively express the QP dispersion as $Z_{0}({\bf k})\varepsilon
_{{\bf k}}$. In the range of $0.01\leq J\leq 0.5$, $Z_{0}({\bf k})\simeq
1.3J/t$. Therefore, $W$ scales with $J$.

Second, we consider the case of strong static JT distortions. The
contribution of static JT interaction is adding an Ising-like
component to the excitations and inducing a large gap in the
orbital excitation spectrum. Thus, the JT effect stabilizes the
orbital ordering. As a result, even for
small $J$, orbital excitations are difficultly stimulated. In fact, for $A_{%
{\bf q}}\gg \xi (\tau )t$ due to either large $E_{{\rm JT}}$ or large $E_{%
{\rm br}}$, Eq. (\ref{se}) can be solved analytically in perturbation
theory: $E_{{\bf k}}\simeq -\xi (\tau )t\gamma _{{\bf k}}-O(\xi ^{2}t^{2}/A_{%
{\bf q}})$. Thus, the QP band is narrowed mainly by latttice
polarons instead of orbital polarons. The QP spectral weight is
\begin{figure}[tbp]
\begin{center}
\epsfig{file=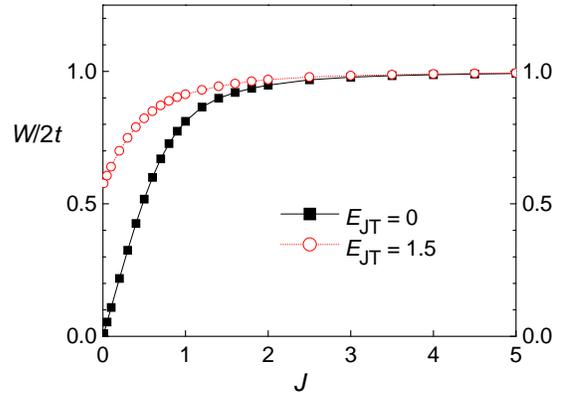, width=\hsize}
\end{center}
\caption{QP bandwidth $W$ as a function of $J$ at $E_{{\rm
br}}=0$.} \label{fig. 3}
\end{figure}
\begin{figure}[tbp]
\begin{center}
\vskip 0.1cm \epsfig{file=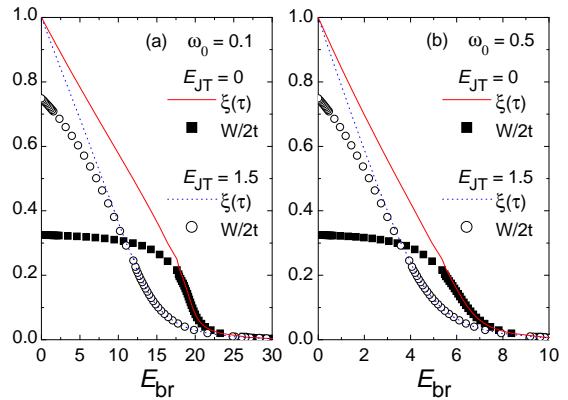, width=\hsize}
\end{center}
\caption{The variation of bandwidth $W$ and the polaronic narrowing $%
\xi(\tau)$ with $E_{{\rm br}}$ at $J=0.3$ for (a) $\omega=0.1$ and
(b) $\omega=0.5$. } \label{fig. 4} 
\end{figure}
\noindent $\xi (\tau )-O(\xi^{2}t^{2}/A_{{\bf q}}^{2})$
approximately. The
QP bandwidth scales with $%
t$. As illustrated in {\rm Fig. 3}, $W\geq 1.5t$ is a slowly
changing function of $J$ at $E_{{\rm JT}}=1.5$. {\rm Fig. 4(a)}
shows $W$ and $\xi (\tau )$ as a function of $E_{{\rm br}}$ at
$E_{{\rm JT}}=1.5$ and $\omega _{0}=0.1$; they behave similarly in
general, indicating that lattice polarization overwhelms orbital
polarization in a wide range of $E_{{\rm br}}$. For not too large
$E_{{\rm br}}$, the small deviation of $W/2t$ from $\xi (\tau )$
is attributed to the remnant hole-orbital-wave coupling. $W$
decreases exponentially with $E_{{\rm br}}$ for $E_{{\rm
br}}>12.5$. This is the Lang-Firsov band narrowing. However, for
the realistic value of $E_{{\rm br}}<5$, $W>1.2t$. Therefore, in
the presence of strong static JT distortions, the QP bandwidth{\em \ }is{\em \ }%
strikingly{\em \ broadened} in comparison with $W\simeq 2.2J$ obtained in
the purely orbital $t$-$J$ model even when the quantum effect of polaronic
{\em band narrowing} is taken into account.

Third, it is interesting to examine the case of no static JT
distortions (i.e., $E_{{\rm JT}}=0$), which is relevant to the
experimental fact that
the static JT distortion rapidly disappears around $x\sim 0.1$ in La$_{1-x}$%
Sr$_{x}$MnO$_{3}$ \cite{Kawano}. For not too large $E_{{\rm br}}$, there is
mix of polarization of phonons and orbital-waves. {\rm Fig. 4(a)} shows the
QP bandwidth $W$ as a function of $E_{{\rm br}}$ at $\omega _{0}=0.1$. For $%
E_{{\rm JT}}=0$, $W/2t$ remains almost unchanged ($\sim 1.1J$) in the range
of $0\leq E_{{\rm br}}\leq 10$ and obviously deviates from the polaronic
band narrowing $\xi (\tau )$ up to $E_{{\rm br}}\simeq 17$. This implies
that the QP is of orbital-polaron type for $E_{{\rm br}}<10$ and is of
polaron type for $E_{{\rm br}}>17$ as well as a mix of the two types for $%
10<E_{{\rm br}}<17$. Therefore, for realistic value of $E_{{\rm br}}<5$, the
electron-phonon interaction effect on the QP band can be neglected. The
hole-orbital-wave scattering dominates the formation of the QP.

Finally, it should be made clear that the small value of $\omega
_{0}=0.1$ (which is however the highest value of phonon frequency
observed in Raman spectra \cite {Iliev}) underlies the
unimportance of quantum effects of electron-phonon interaction on
the QP band. Note that the lattice polaron effect is the
most pronounced in the antiadiabatic limit ($\omega _{0}\rightarrow \infty $%
), where $\gamma \rightarrow 1$, and is negligible in the adiabatic limit ($%
\omega _{0}\rightarrow 0$), where $\gamma \rightarrow 0$. At $\omega
_{0}=0.1 $, $\tau <0.1$, the system is close to adiabatic limit. {\rm Fig.
4(b)} shows the same quantities as {\rm Fig 4(a)} but at $\omega _{0}=0.5$
which is chosen in Ref. \cite{Roder}. The quantum effects of hole-phonon
interaction are improved remarkably. For example, the critical values of $E_{%
{\rm br}}$ at which different QP types occur are 3 times as smaller as those
at $\omega _{0}=0.1$.

Summarizing, we present a systematic study on single hole motion
in LaMnO$_{3}$. We show that in the realistic parameter space for
LaMnO$_{3}$, quantum effects of hole-phonon interaction are small.
In the purely orbital $t$-$J$ model,
the QP bandwidth $W$ scales with $J$. Considering the hole-phonon coupling, $%
W$, which however scales with $t$, is strikingly broadened by strong static
JT lattice distortions even when the polaronic band narrowing is taken into
account. We predict that orbital polarization is pronounced for $x>0.1$
where static JT distortions disappear. Our results can be tested by future
angle-resolved photoemission spectroscopy (ARPES) experiments.

The authors thank P. W. Leung, F. C. Zhang, and H. Zheng for useful
discussions. The ED program was implemented on the basis of $C$ codes
provided by P. W. Leung. W.G.Y. is supported by CUHK 4288/00P 2160148.

{\em Note added.---}Very recently, two papers concerning ARPES of LaMnO$_{3}$
have appeared. Perebeinos and Allen \cite{Perebeinos} addressed the
Frank-Condon broadening effect driven by electron-phonon
interaction on the bare hole dispersion in the JT-ordered ground state.
Brink, Horsch, and Ole\'{s} (BHO) \cite{Brink-new} calculated the
single-hole spectral functions for the purely orbital $t$-$J$ model using
the SCBA and discussed the crystal-field effect on the QP band. Their pieces
of work are in a sense complementary to ours.
The present work takes into account both of orbital polarization and
lattice polarization, giving rise
to a different, yet more realistic picture from BHO's.




\begin{references}
\bibitem{note1}  We have assumed here that the JT modes of lattice
distortions can be treated classically. At low temperature static JT
distortions are so strong that the classical approximation is appropriate as
well as quantum JT effects can be be absorbed into the breathing mode \cite
{Millis,Roder}.
\end{references}
\end{document}